\documentstyle[preprint,eqsecnum,aps]{revtex}
\def\lsim{\:\raisebox{-0.5ex}{$\stackrel{\textstyle<}{\sim}$}\:}
\def\gsim{\:\raisebox{-0.5ex}{$\stackrel{\textstyle>}{\sim}$}\:}
\begin{document} 
\draft 

\title{Statistical Model for the Nucleon Structure Functions}

\author{R.S. Bhalerao$^*$}
\address{Theoretical Physics Group\\
Tata Institute of Fundamental Research\\
Homi Bhabha Road, Colaba, Bombay 400 005, India}

\maketitle

\begin{abstract}
A phenomenological model for the nucleon structure functions is
presented. Visualising the nucleon as a cavity filled with parton gas
in equilibrium and parametrizing the effects due
to the finiteness of the nucleon volume, we obtain a good fit to the
data on the structure function $F_2^p$. The model then successfully
predicts other unpolarized structure function data.
\end{abstract}



\vfill
\noindent{$^*$E-mail: bhalerao@theory.tifr.res.in, ~~ Fax: 091 22 215 2110}

\newpage
Recent experiments have revealed some remarkable features of the
nucleon structure functions $F_2^{p,n}$. Data on deep inelastic
scattering of muons off proton and deuteron targets \cite{nmcgsr}
show that the quark sea in the nucleon is not flavor-symmetric,
$\bar u(x) \neq \bar d(x)$; the Gottfried sum \cite{gsr} $S_G \equiv
\int \left( F^p_2 - F^n_2 \right) (dx/x)$, at $Q^2 = 4$ GeV$^2$,
has the value $0.235 \pm 0.026$ compared to the usual quark model
prediction of 1/3. This result has been confirmed by the observed
asymmetry in Drell-Yan production of dileptons in $pp$ and $pn$
collisions \cite{NA51}. Most
notably, the HERA electron-proton scattering data \cite{hera} reveal a
rapid rise of the proton structure function $F^p_2 (x)$ as $x$
decreases. Indeed over a wide range of small $x$, data from the
various groups \cite{hera,nmcbcdms}, for fixed $Q^2$,
are all well described by a single
inverse power of $x$. Figure 1 is a log-log plot of the data on
$F_2^p(x)/x$ (the combination that enters $S_G$) versus $x$. We see
that, for fixed $Q^2$, the data fall on straight lines defined by
\begin{equation}
{F^p_2(x) \over x}={c\over x^m}~, ~(0.0004 \lsim x \lsim 0.2).
\eqnum{1}
\end{equation}
For instance, at $Q^2=15$ GeV$^2$, the best-fit parameters are
$c=0.229\pm 0.005$ and $m= 1.22 \pm 0.01$.\footnote{For $Q^2 = 35,
120$ GeV$^2$ the power-law fits give, respectively, $c=0.229\pm
0.006,~0.163\pm 0.037$ and $m=1.26\pm 0.01,~1.35\pm 0.04 $.}
{\it In Fig. 1, the straight-line extrapolations in the unexplored low-$x$
region are our predictions based on Eq. (1).}

Global fits to the nucleon structure data involve parametrizing the
various parton distributions at some low $Q^2$ and evolving them to
higher $Q^2$ relevant to observations. The fits so obtained
\cite{params} have very high precision but contain several (typically
$\sim $15-20) arbitrary parameters and provide little physical insight
into the structure of the nucleon. On the other hand, phenomenological
models could give us some valuable clues into the physics of parton
distributions in the nucleon. From this point of view 
the parton gas models or the statistical
models of the structure functions \cite{gasms} have been quite
interesting due to their intuitive appeal and simplicity. {\it Bickerstaff
and Londergan \cite{bl} have provided a strong justification for the
general philosophy of the statistical models and have also discussed
limitations of other models.}

We present here a phenomenological model for the unpolarized nucleon
structure functions by using ideas from statistical mechanics. As a
starting point, we assume that inside the nucleon, the valence quarks
together with the sea quarks, antiquarks and gluons constitute a
noninteracting gas in equilibrium. (It may be
recalled that particles are treated as noninteracting even in the
quark-parton model of Feynman.) This simple picture is then improved
upon in two respects: (i) Finite-size (of the nucleon) corrections
(FSC) to the statistical expression for the number of states per unit
energy interval are taken into account, and (ii) the resulting
structure functions are evolved to the experimental values of $Q^2$ by
using the standard techniques in 
quantum chromodynamics (QCD). Each of these two effects is
shown to play an essential role. To our knowledge, calculation of the
structure functions taking into account FSC has never been reported in
the literature. The model reproduces all unpolarized structure
function data from $x \simeq 1$ to $x \simeq 10^{-4}$ quite well.

\medskip

\noindent {\it The Model} 

We picture the nucleon (mass $M$) to consist of a gas of massless
partons (quarks, antiquarks and gluons) in equilibrium at
temperature $T$ in a spherical volume $V$ with radius $R$. We consider
two frames, the proton rest frame and the infinite-momentum frame
(IMF) moving with velocity $-v (\simeq -1)$ along the common $z$
axis. Our interest lies in the limit when the Lorentz factor $\gamma
\equiv (1-v^2)^{-1/2} \rightarrow \infty$. The invariant parton number
density in phase space \cite{degroot} is given by (quantities in the
IMF are denoted by the index $i$)
\begin{equation}
{dn^i \over d^3p^i~d^3r^i}={dn\over d^3p~d^3r} 
= {g f(E) \over (2\pi )^3}~, \eqnum{2} 
\end{equation}
where $g$ is the degeneracy ($g=16$ for gluons and $g=6$ for $q$ or
$\bar q$ of a given flavor), $(E,\bf p)$ is the parton four-momentum
and $f(E) = \{\exp[\beta (E-\mu)]\pm 1\}^{-1}$ is the usual
Fermi or Bose distribution function with $\beta \equiv T^{-1}$. In
order to obtain the number distribution $dn^i/dx$ in the Bjorken scaling
variable $x =p^i_z/(Mv \gamma)$, we note that
\[
d^3p^i~d^3r^i = 2\pi p^i_Tdp^i_T (Mv \gamma dx) d^3r / \gamma
= 2 \pi M^2 x dE~dx~d^3r,
\]
and
\[
d^3p~d^3r = 4 \pi E^2~dE d^3r,
\]
for massless partons. For fixed $x$ the parton energy $E$ varies
between the kinematic limits $(xM/2)$ and $(M/2)$, where the lower
limit is attained when $p^i_T = 0$ and the upper limit follows simply
from energy-momentum considerations. (The kinematics will be described
in detail in \cite{rsbkvl}.) Hence $dn^i/dx$ in the IMF is related to
$dn/dE$ and $f(E)$ in the proton {\it rest frame} as follows
\begin{equation}
{dn^i \over dx} 
= {M^2 x \over 2} \int ^{M/2}_{xM/2} {dE \over E^2} {dn \over dE}
= {g V M^2 x \over (2\pi)^2} \int ^{M/2}_{xM/2} dE f(E).\eqnum{3} 
\end{equation}
The structure function $F_2(x)$ is given by
\[
F_2(x) = x \sum_q e_q^2 \bigg\{ 
\left({dn^i \over dx}\right)_q + \left({dn^i \over dx}\right)_{\bar q}
\bigg\}.
\]
The number distribution $dn^i/dx$ in Eq. (3) 
vanishes linearly as $x \rightarrow 0$ (and
also as $x \rightarrow 1$) and leads to the behavior of the structure
function $F_2(x) \sim x^2$ at small $x$, which disagrees with the
observations noted in Eq. (1).
  
In an attempt to obtain the rise of $F_2^p(x)$ at small $x$, we now
explore the effects arising from the finiteness of the nucleon volume
$V$. We note from Eq. (2) that $dn/dE = g f(E) V E^2 /2\pi^2$, which
is strictly valid in the large-volume limit, i.e., when the surface
and curvature terms are negligible. We shall modify the model by
incorporating these subleading terms. Various studies \cite{fse} of 
finite-size corrections (FSC) show that they depend on the particular
equation of motion that is employed and
are sensitive to the precise shape
and size of the enclosure, the type of boundary conditions imposed on
the wave function, and to the details such as whether the particles
are strictly massless. Moreover, these studies invariably
involve some simplifying assumptions and thus a blind use of their
results cannot be justified in the present context. We feel more work
needs to be done to fully understand the finite-size effects in a QCD
bound state such as the nucleon.

In keeping with the phenomenological nature of the model, we have
incorporated the FSC by rewriting $dn/dE$ in Eq. (3) as in \cite{fse}:
\begin{equation} 
dn/dE = g f(E) (VE^2/2\pi^2 + aR^2E + bR),
\eqnum{4}
\end{equation}
and treating the numerical coefficients $a$ and $b$ in the surface
and curvature terms as free parameters for reasons stated in the
previous paragraph.

The model described above is assumed to hold at a certain input momentum
scale $Q_0^2$, and if necessary can be evolved to higher $Q^2$ by means
of the standard techniques in QCD.
To complete the statement of the model, we require the parton
distributions to obey the following {\it three} constraints
at the input scale. The
constraints on the net quark numbers in the proton are $n_u - n_{\bar u} =
2$ and $n_d - n_{\bar d} = 1$, i.e.,
\begin{equation}
{M^2 \over 2} \int ^1_0 dx~ x~\int^{M/2}_{xM/2}{dE \over E^2}~
\bigg\{ \left({dn \over dE}\right)_\alpha -
 \left({dn \over dE}\right)_{\bar\alpha} \bigg\} 
= n_\alpha -n_{\bar\alpha}.~~~~~~~~~~(\alpha = u,d) \eqnum{5} 
\end{equation}
Obviously, chemical potentials for heavy flavors are necessarily zero.
As regards the third constraint, we assume that the
longitudinal momentum fractions in the $u,~d$ flavors and the gluons
add up to unity:
\begin{eqnarray}
{M^2 \over 2} \int ^1_0 dx~x^2 \int^{M/2}_{xM/2} {dE \over E^2}~
\bigg\{ 
  \left({dn \over dE}\right)_u 
+ \left({dn \over dE}\right)_{\bar u}
+ \left({dn \over dE}\right)_d
+ \left({dn \over dE}\right)_{\bar d} 
+ \left({dn \over dE}\right)_g \bigg\}  = 1~. \eqnum{6} 
\end{eqnarray}
The quark flavors $s$ and $c$ which are not introduced in Eq. (6)
show up at higher $Q^2$ as a result of QCD evolution. 

By interchanging the order of $x$ and $E$ integrations in Eqs. (5-6)
and performing the $x$-integration analytically, we see that in order
to keep the integrals finite, large powers of $1/E$ are not allowed in
the integrand. This means that the three terms 
displayed in Eq. (4) are the only ones that are allowed.
{\it Thus the model effectively has only two free parameters
$a$ and $b$.}

The temperature ($T$) and
two chemical potentials ($\mu_u,~ \mu_d$) are not free parameters;
we determine them by solving the three coupled
nonlinear equations (5-6) by the Davidenko-Broyden method
\cite{antia}. The resulting values of $T$, $\mu_u$ and $\mu_d$ are 
such that the left and right hand sides of these equations agree with
each other to typically one part in $10^6$. 
The parton distributions were evolved by means of the
Dokshitzer-Gribov-Lipatov-Altarelli-Parisi equations \cite{glap} 
in leading order,
taking the input scale $Q^2_0=M^2$ and $\Lambda_{QCD}=0.3$ GeV.
Finally, the root-mean-square (rms) radius of the parton
distribution was taken to be the same as the charge rms radius
($\rho$) of the proton; since $\rho \simeq 0.862$ fm \cite{simon},
this yields $R = \sqrt{5/3}~ \rho = 1.11$ fm.

\medskip
\noindent {\it Results and Discussion}

Since the two arbitrary constants $a$ and $b$ in
Eq. (4)  are not known, we have determined them by fitting the
deep inelastic scattering data on $F^p_2(x)$ at $Q^2=15$ GeV$^2$
\cite{hera,nmcbcdms}. The results of our fit incorporating
QCD evolution and FSC are shown by the solid curve in
Fig. 2. Also shown for comparison in
Fig. 2 are: (a) the (dot-dashed) curve labeled `GAS' giving the
prediction of the (unmodified) parton gas model which has no free 
parameters by
virtue of the constraints, (b) the (dashed) curve labeled `QCD'
showing the effect of QCD evolution on the gas model, and (c) the
(dotted) curve labeled `FSC' showing a {\it fit} to the data when only
the FSC are introduced in the gas model. For the solid curve in Fig. 2,
the fitted values of the two parameters are
$a = -0.400$ and $b = 0.475$,
and the corresponding temperature and chemical potentials are $T
= 72$ MeV, $\mu_u = 162$ MeV and $\mu_d = 81$ MeV.\footnote{It is 
amusing to note that the values of $a$
and $b$ determined by us are close to the values
$a=-1/2$ and $b=3/(2 \pi)$ which follow from one of the expressions for
$dn/dE$ given by Morse and Ingard \cite{fse}; substitute $c = 1, \nu
= E/(2\pi), A=4 \pi R^2 $ and $\ell_x =\ell_y=\ell_z=2R$ in $dN_{ob}$
in their Eq. (9.5.12).}

As a test of the model, we show in Fig. 3 the prediction (solid curve)
for the difference $[F^p_2(x) - F^n_2(x)]$. Also shown for comparison
is the result (dashed curve) based on the parametrization of Gl\"uck
et al. \cite{params}. The agreement with the NMC data is
reasonable.

As for the Gottfried sum $S_G$, we have 
\[
S_G = {1 \over 3} - {2 \over 3}~{M^2 \over 2} \int^1_0 dx ~x
\int^{M/2}_{xM/2} {dE\over E^2}~ 
\bigg\{\left({dn\over dE}\right)_{\bar d} - 
       \left({dn\over dE}\right)_{\bar u}\bigg\}.
\]
\noindent The inequality $S_G < {1 \over 3}$ is thus a result of
having in the proton, more valence $u$ quarks than valence $d$ quarks,
$(n_u - n_{\bar u}) > (n_d - n_{\bar d})$, implying that $\mu_u >
\mu_d$ (see Eq. (5))
and hence the integral in $S_G$ is positive. Our model
predicts at $Q^2 =4$ GeV$^2$, the value $S_G=0.22$ which is
consistent with the experimental value $S_G=0.235\pm 0.026$.
{\it The natural explanation of the violation of the Gottfried sum rule
is an attractive feature of our general framework.}

In addition, the model is in excellent agreement with data on the
gluon distribution $g(x)$. It also reproduces well the ratio
$F^n_2(x)/ F^p_2(x)$, the ratio $\bar u(x)/ \bar d(x)$ at $<x>=0.18$
which has been deduced to be about $0.51$ by the NA51 collaboration
\cite{NA51}, the longitudinal momentum fraction carried by
the charged partons, etc. These and other predictions of the model, on
the quark and antiquark distributions $q(x)$, $\bar q(x)$, $q_v(x) =
q(x) - \bar q(x)$ for various flavors, etc. will be discussed in
detail elsewhere \cite{rsbkvl}.

Now we briefly describe the salient features of some of the recent
calculations of the nucleon structure functions, which use ideas from
statistical mechanics. It has generally been difficult so far to model
$F_2^p(x,Q^2)$ such that it is nonzero only in $x=[0,1]$, and is
consistent with the two number constraints, the momentum constraint,
the Gottfried sum and the observed low-$x$ rise.
Mac and Ugaz [7a] calculated first-order QCD corrections to the
statistical distributions and obtained a crude but reasonable
agreement with $F_2^p(x)$ data for $x \gsim 0.2$. The momentum
constraint was not imposed and the fitted value of the proton radius
was 2.6 fm.
Cleymans et al. [7b] used the framework of the finite
temperature quantum field theory. They considered ${\cal O}(\alpha_s)$
corrections to the statistical distributions and obtained a good fit
to the $F_2^p(x)$ data for $x \geq 0.25$. They also calculated the
ratio $\sigma_L / \sigma_T$ in this region; it was a factor of 6 above
the experimental value.
Bourrely et al. [7c] considered polarized as well as
unpolarized structure functions and presented a statistical
parametrization (with eight parameters) of parton distributions in the
IMF. Their framework allowed chemical potential for quarks as well as for
gluons. The number constraints were not satisfied very accurately. QCD
effects were not considered. Parton distributions were nonzero for
$x > 1$. $x \bar q(x)$ vanished as $x \rightarrow
0$ and so it was not possible to reproduce the fast increase of the
antiquark distributions for $x<0.1$.
Bourrely and Soffer's [7d] approach was similar to that in [7c]. By
incorporating QCD evolution of parton distributions and allowing the
antiquark chemical potential to depend on $x$, they were able to
reproduce the HERA data on $F_2^p$.
Recently Buccella et al. [7e] have introduced in this model a so-called
liquid term to reproduce the low-$x$ behaviour of structure functions.

In conclusion, it is noteworthy that the application of ideas of
statistical mechanics to the point constituents of the nucleon can
provide a simple description of all the observed features of the
unpolarized nucleon structure functions down to the lowest $x$
values so far explored. The model presented here has two free
parameters which arise from our treatment of the finite-size
corrections. It is hoped that the success of the model would provide
a stimulus to further studies of the finite-size effects in a
QCD bound state such as the nucleon.

\acknowledgments

I am grateful to K.V.L. Sarma for numerous discussions during the
course of this work. I thank R.M. Godbole and S. Kumano for
discussions and communications regarding QCD evolution. I benefited
from valuable comments by A.K. Rajagopal, Virendra Singh and
C.S. Warke.

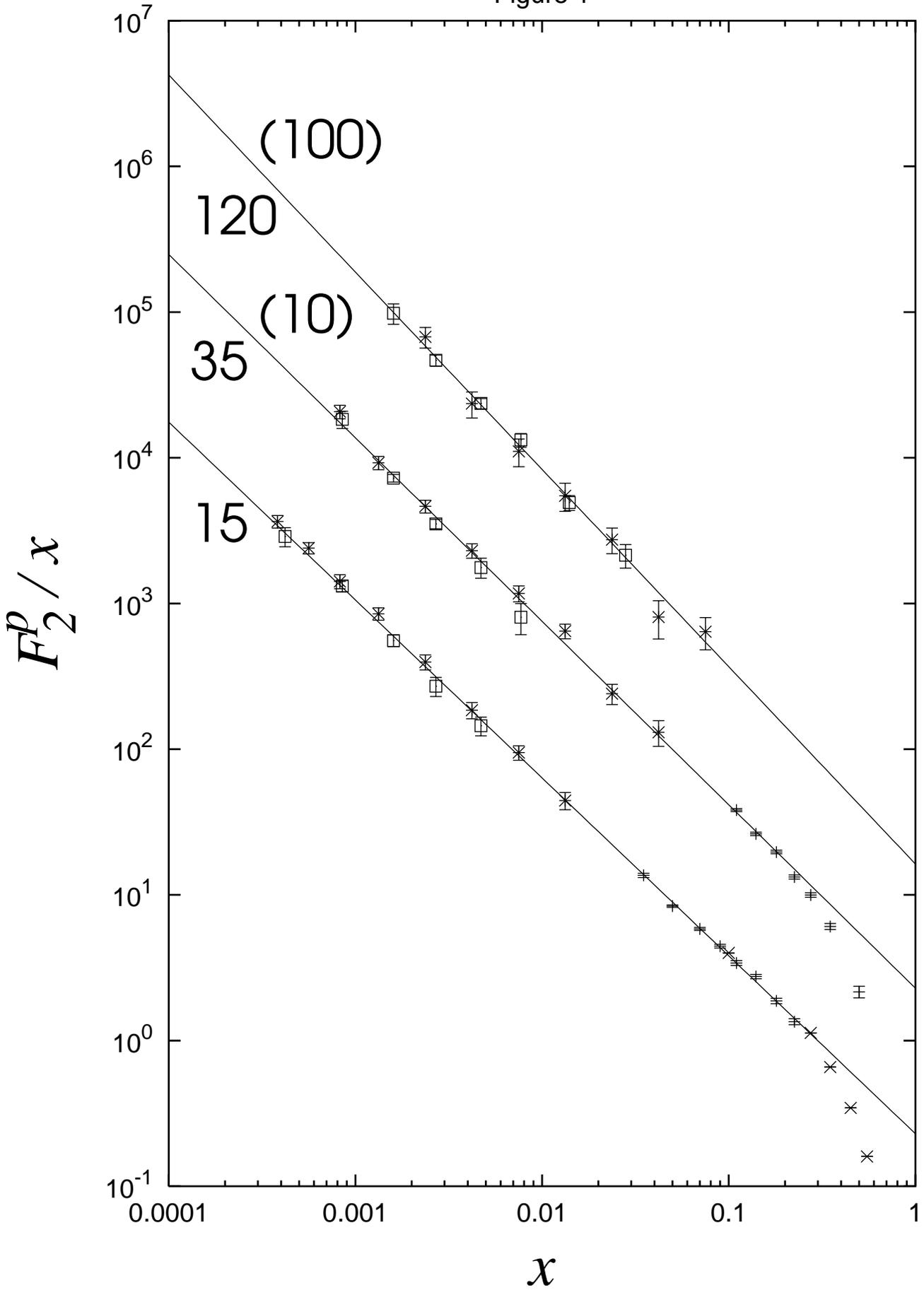
\begin{figure} 
\caption
{Log-log plot of the proton structure function data. 
Experimental data are from
Refs. [4,5]; the error bars show statistical and
systematic errors combined in quadrature. The straight lines
are our fits described in Eq. (1), and are labeled by $Q^2$ = 15,~35,
and 120 GeV$^2$. Numbers have been scaled by the factors shown in
parentheses for convenience in plotting.}
\label{1}
\end{figure}

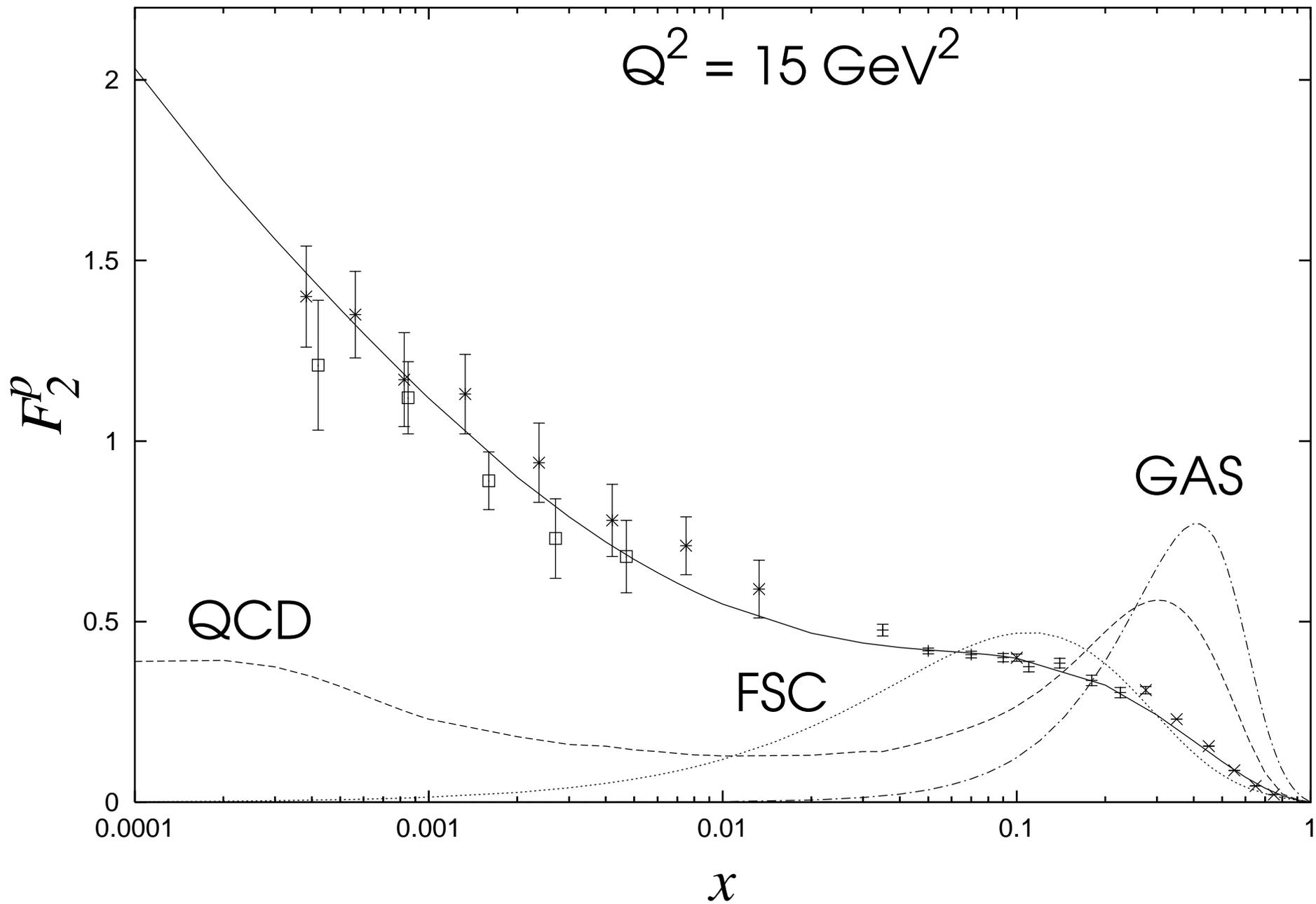
\begin{figure} 
\caption
{Proton structure function $F^p_2(x)$ at $Q^2 = 15$ GeV$^2$. Data points
are as in Fig. 1. 
Solid curve is our best fit to the data and includes both FSC and QCD.
Also shown for comparison
are: the (dot-dashed) curve labeled `GAS' giving the (unmodified) gas model
prediction, the (dashed) curve labeled `QCD' showing the QCD-evolved
gas model, and the (dotted) curve labeled `FSC' which is a fit to the
data when finite-size corrections are included in the gas model
(without QCD).}
\label{2}
\end{figure}

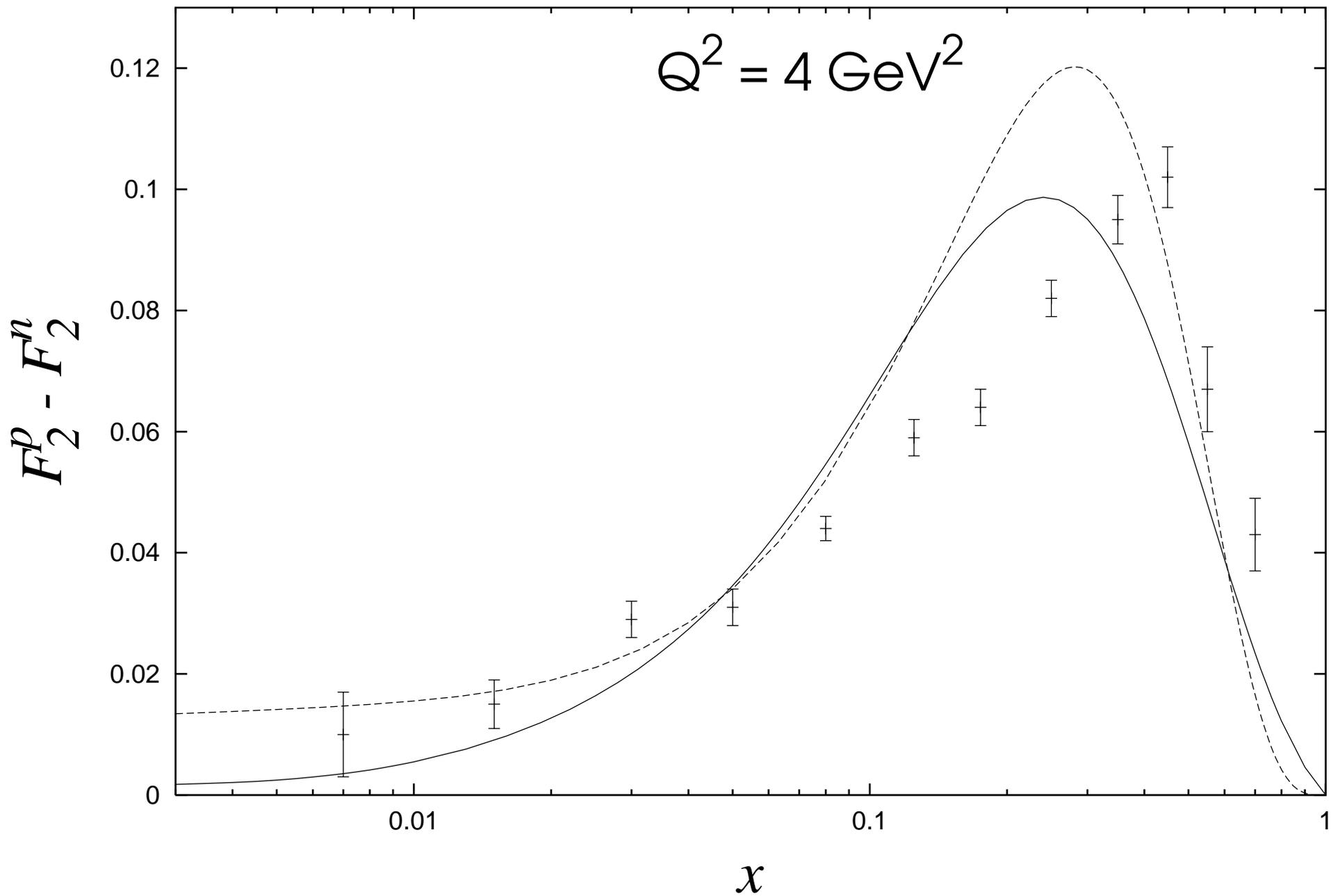
\begin{figure} 
\caption
{Difference $(F^p_2-F^n_2)$ versus $x$, at $Q^2
= 4$ GeV$^2$. Experimental data are from 
Ref. [1];
errors are statistical only. Solid curve is the prediction of our model.
Dashed curve is based on the parametrization of Gl\"uck et al. [6].}
\label{3}
\end{figure}

\end{document}